\documentclass[sigconf]{acmart}

\AtBeginDocument{%
  \providecommand\BibTeX{{%
    \normalfont B\kern-0.5em{\scshape i\kern-0.25em b}\kern-0.8em\TeX}}}

\acmConference[Autodesk Research]{San Francisco}{San Francisco}{USA}{}
\acmBooktitle{Autodesk Research}{}{}

\usepackage[T1]{fontenc}
\usepackage[utf8]{inputenc}
\begin{document}

\title{SketchOpt: Sketch-based Parametric Model Retrieval for Generative Design}

\author{Mohammad Keshavarzi}
\orcid{1234-5678-9012}
\affiliation{%
  \institution{University of California, Berkeley}
  \streetaddress{337 Cory Hall}
  \state{CA}
  \country{United States}}
\email{mkeshavarzi@berkeley.edu}

\author{Clayton Hutson}
\affiliation{%
  \institution{Autodesk Research}
  \city{San Francisco, CA}
  \country{United States}}
\email{clayton.hutson@autodesk.com}

\author{Chin-Yi Cheng}
\affiliation{%
  \institution{Autodesk Research}
  \city{San Francisco, CA}
  \country{United States}}
\email{chin-yi.cheng@autodesk.com}

\author{Mehdi Nourbakhsh}
\affiliation{%
  \institution{Autodesk Research}
  \city{San Francisco, CA}
  \country{United States}}
\email{mehdi.nourbakhsh@autodesk.com}

\author{Michael Bergin}
\affiliation{%
  \institution{Autodesk Research}
  \city{San Francisco, CA}
  \country{United States}}
\email{michael.bergin@autodesk.com}

\author{Mohammad Rahmani Asl}
\affiliation{%
  \institution{Autodesk Inc.}
  \city{San Francisco, CA}
  \country{United States}}
\email{mohammad.asl@autodesk.com}

\renewcommand{\shortauthors}{Keshavarzi et al.}

\begin{abstract}
Developing fully parametric building models for performance-based generative design tasks often requires proficiency in many advanced 3D modeling and visual programming, limiting its use for many building designers. Moreover, iterations of such models can be time-consuming tasks and sometimes limiting, as major changes in the layout design may result in remodeling the entire parametric definition. To address these challenges, we introduce a novel automated generative design system, which takes a basic floor plan sketch as an input and provides a parametric model prepared for multi-objective building optimization as output. Furthermore, the user-designer can assign various design variables for its desired building elements by using simple annotations in the drawing. The system would recognize the corresponding element and define variable constraints to prepare for a multi-objective optimization problem.
\end{abstract}



\begin{teaserfigure}
  \includegraphics[width=\textwidth]{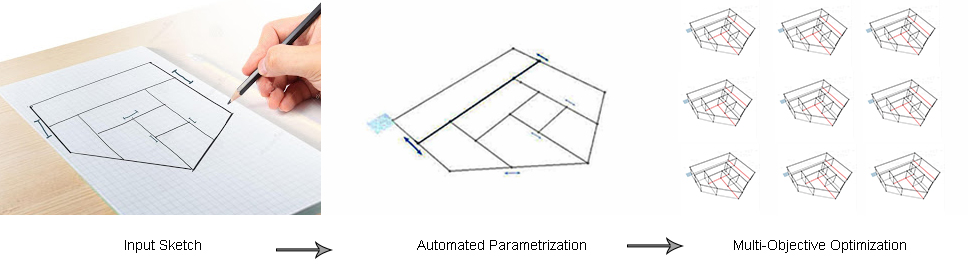}
  \caption{Abstract illustration of the search process a) stochastic search b) Recommendation system c) Hybrid approach.}
  \label{fig:teaser}
\end{teaserfigure}

\maketitle

\begin{figure*}
\centering
  \includegraphics[width=1.95\columnwidth]{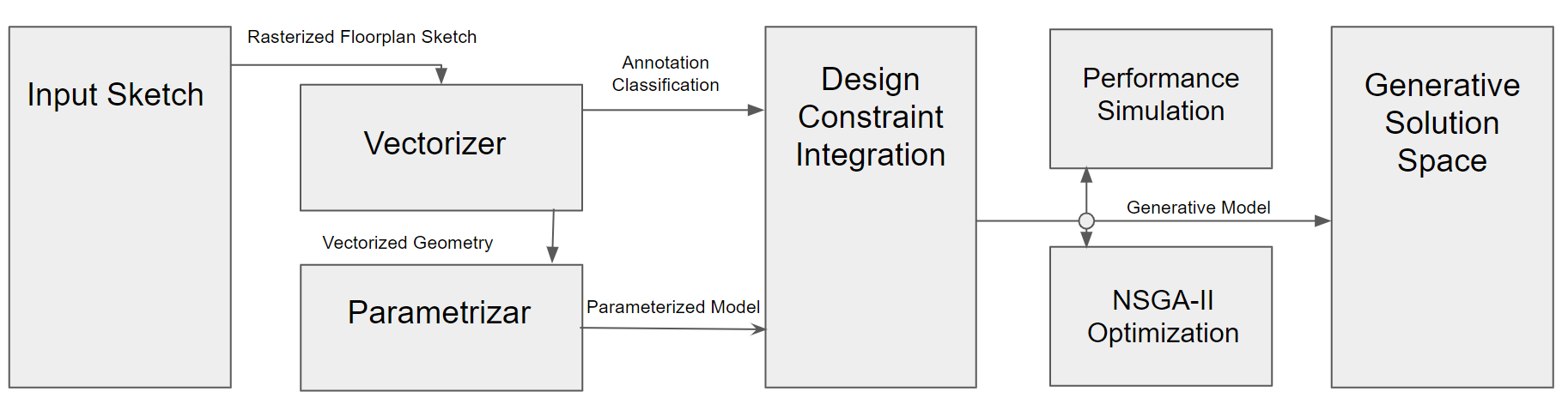}
  \caption{Abstract illustration of the search process a) stochastic search b) Recommendation system c) Hybrid approach.}~\label{fig:conceptFig}
\end{figure*}

\section{Introduction}
In recent years, performance-based generative design systems (PGDS) have been widely adopted in the building design process, altering iterative trial-and-error cycles, to efficient optimization-based workflows. Using such systems, the user first programmatically defines the design parameters and constraints and then establishes desired building performance goals. This would allow the system to search for a solution that best matches the target performance criteria in the generated solution space. PDGS have been widely deployed in visual programming tools, such as Grasshopper3D \cite{rutten2015grasshopper3d} and Autodesk Dynamo, where they can efficiently integrate with building performance simulation software and optimization algorithms. However, this process often requires a spectrum of technical expertise in the modeling and parametrization stages, excluding many designers from taking advantage of such an approach. Furthermore, modifying the parametric logic and constraints of generative models can be time-consuming tasks, limiting the ideal iterative process between the designer and the system. 

To address this challenge, we introduce SketchOpt, an interactive sketch-based generative design system for early-stage building performance optimization tasks. Unlike conventional PGDS workflows where design definitions, constraints, and objective functions are defined in a programmatic fashion, SketchOpt takes hand-drawn floorplans and annotations as input and outputs fully parametric models prepared to integrate with building performance simulation software and multi-objective optimization algorithms.  Our approach takes advantage of machine learning and procedural modeling methodologies to seamlessly merge interactive sketching to a performance-based generative design process. This would enable designers to take advantage of the intuitiveness, freedom, and flexibility of sketching while leveraging the search and optimization potentials offered in a parametric model. The user does not need to program tedious parametric definitions; instead, they are recognized from the sketch and annotations in a real-time fashion, enabling untrained users to quickly create complex parametric models

\section{Related Work}

One of the main components of our system is the ability to convert a sketch-based rasterized input to a vectorized parametrized model. A common methodology that has been widely explored in previous studies is to use various matching strategies to compare the similarity of input sketches to an existing database of 2D or 3D models \cite{eitz2012sketch, funkhouser2003search, hou2006sketch, schneider2014sketch}. However, these systems only allow retrieval of existing 3D models and provide no means to create new 3D models. More recently, sketch-based shape retrieval has been demonstrated through convolutional neural networks (CNNs). CNNs are able to learn hierarchical image representations optimized for image processing performance. \cite{krizhevsky25hinton}. Su et al. \cite{su2015multi} performed sketch-based shape retrieval by adopting a CNN architecture pre-trained on images, then fine-tuning it on a dataset of sketches collected by human volunteers \cite{eitz2012humans}. Wang et al. \cite{wang2015sketch} used a Siamese CNN architecture to learn a similarity metric to compare human and computer-generated line drawings. Abbasi-Asl et al. take advantage of Brain-Computer Interface systems for sketch-based model retrieval in virtual environments \cite{abbasi2019brain}. Nishida et al. \cite{nishida2016interactive} introduced a CNN-based urban procedural model generation from sketches. However, instead of solving directly for the final shape, their method suggests potentially incomplete parts that require further user input to produce the final shape. In contrast, our approach is end-to-end, requiring users to provide only an approximate sketch of the floorplan layout.

In the field of performance-based generative design, previous research can be initially categorized based on the performance criteria itself. Works of Asadi and Geem \cite{asadi2015sustainable} and Attia et al. \cite{attia2013assessing} provides a technical review of relevant research applying simulation-based optimization methods to sustainable building design. In terms of architectural design strategies, Turrin et al. \cite{turrin2011design} discuss the benefits derived from combining parametric modeling and Genetic Algorithms (GAs) to achieve a performance-oriented process in design, with specific focus on architectural design, with a focus on the key role played by geometry in architecture. Huang and Niu \cite{huang2016optimal} analyze the history, current status, and potential of optimal building design based on the simulation performance.

\begin{figure}
\centering
  \includegraphics[width=1\columnwidth]{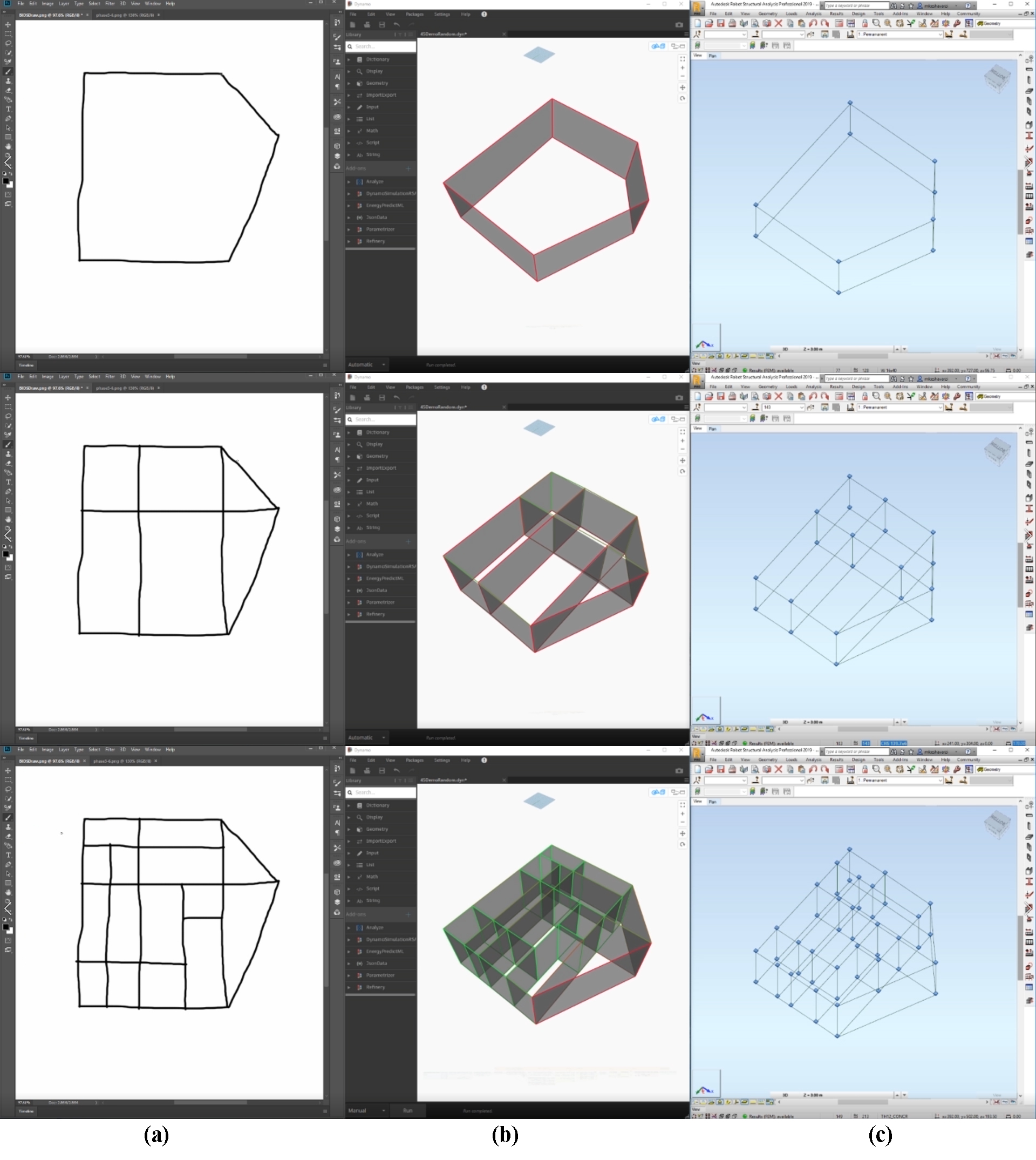}
  \caption{Multiple stages of the sketching process (a) with automated real-time parametrization in Dynamo VPL environment (b) and structural simulation in Autodesk Robot (c)}~\label{fig:sketchPorc}
\end{figure}

Basbagill at al \cite{basbagill2014multi} propose a multi-objective feedback approach for determining life-cycle environmental impact and cost performance of buildings at the conceptual design stage, exploring the concept of multidisciplinary design optimization. Brown and Mueller \cite{brown2016design} also present a multidisciplinary optimization model for structural and energy performance at early stage design phases. In that work, the structural optimization objective was to minimize the amount of steel required, and the energy optimization objective was to reduce the annual energy of the building in terms of lighting, heating, and cooling. The Non-Dominated Sorting Genetic Algorithm II (NSGA-II) \cite{deb2002fast} was used to iteratively approach the Pareto front over several generations of design alternatives of three types of long steel span structures. Keshavarzi et al. \cite{keshavarzi2020optimization} take advantage PGDS workflows to find alternative indoor furniture arrangements for spatial computing telepresence systems. Lin and Gerber \cite{lin2014designing} focus on early-stage design decisions, proposing a framework for concept design based on multiple design objectives, using a custom GA for design and performance optimization. Finally, there is a large body of research focused on how to explore generative design solutions, and to provide the user with appropriate workflows to modify and interact with the generated solution space \cite{scott2002investigating,natureHardy,meignan2015review, keshavarzi2020v, mueller2015combining}.

\section{Methodology}

The SketchOpt system consists of four main components: i) the vectorizer module, where hand-drawn rasterized floorplan sketches are converted to vectorized data structures. ii) the parametrized module, where various building elements are joined and constrained in relation to each other according to their vectorized geometric properties. iii) parsing user-defined optimization constraints and variables, where annotations in the drawing are detected and applied to the parametric model. iv) and finally, integration with the NSGA-II optimization module for customized performance-based optimization tasks.

\begin{figure}
\centering
  \includegraphics[width=1\columnwidth]{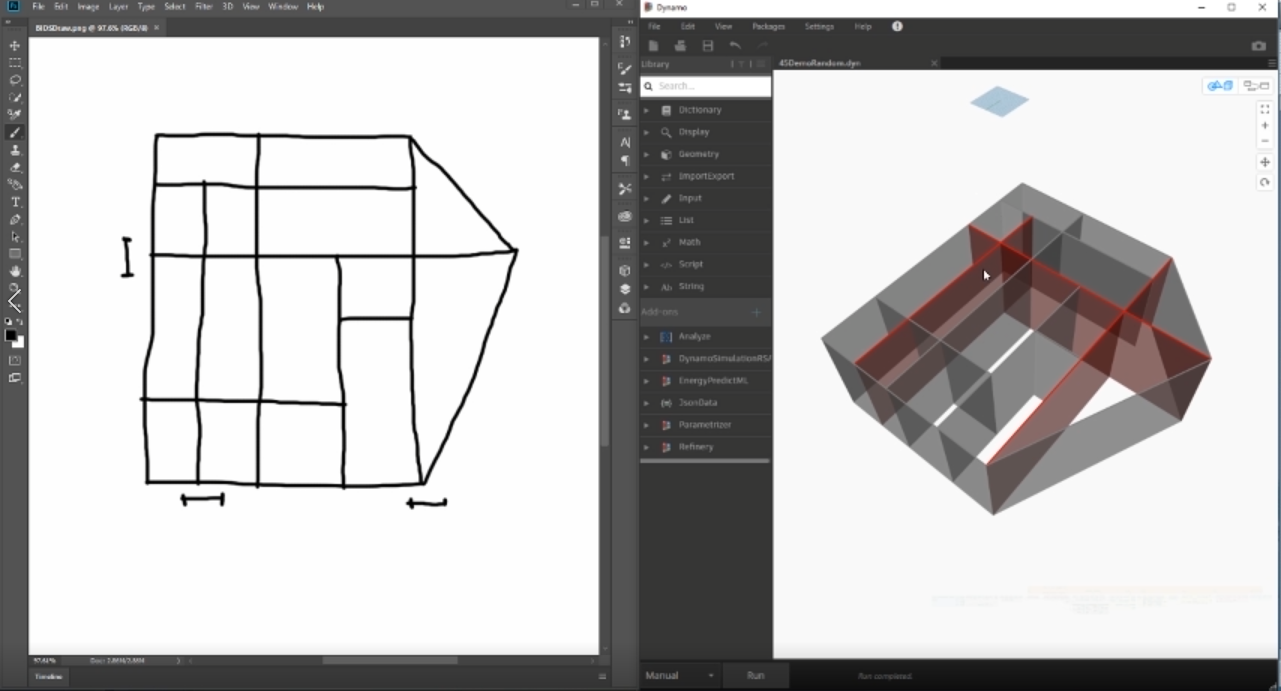}
  \caption{Applying optimization parameters and constraints by sketching annotations}~\label{fig:applyParam}
\end{figure}

\subsection{Vectorizer}

The vectorizer module integrates a special form of asymmetric convolution to detect linear features present at just a few hundredths of a percent of overall luminosity and run at 5 independent resolutions simultaneously.  For each node (pixel) found to contain linearity, the resolutions participate in a weighted optimization to compute the resulting output orientation and gain. The core algorithm is parallelized to run on multi-GPU environments as well as AVX CPU instructions and can vectorize a 3k by 3k scene at several resolutions in a few hundred milliseconds. The vectorizer module is an investigation and prototype of computer vision techniques that support extraction of vector and linear information from raster sources, including photos of plans or hand-drawn sketches. The applied use-case in this project is for architects to create or import conceptual designs in a fast manner with a minimum of errors.

\subsection{Parametrizer}

The Parametrizer algorithm is responsible to apply geometric grammar and topological relationships between the vectorized geometrical data. This is achieved by considering contextual and mutual attributes of each of the vectorized segments and converting the data to a graph system. Building elements such as walls, columns, etc. are automatically generated using predefined architectural assumptions and clustered based on type, position, and continuity. Moreover, all groups are constrained by their neighboring group resulting in joint variable parameters for the generated elements. As each member element is defined by a corresponding node and path, effective translations applied to one element group would also apply to the connected member of the neighboring groups. Such attribute would allow geometric translations of each group to apply in a parametric fashion, with modifications happening in a constrained range while maintaining the overall geometric layout. 

\begin{figure*}
\centering
  \includegraphics[width=2\columnwidth]{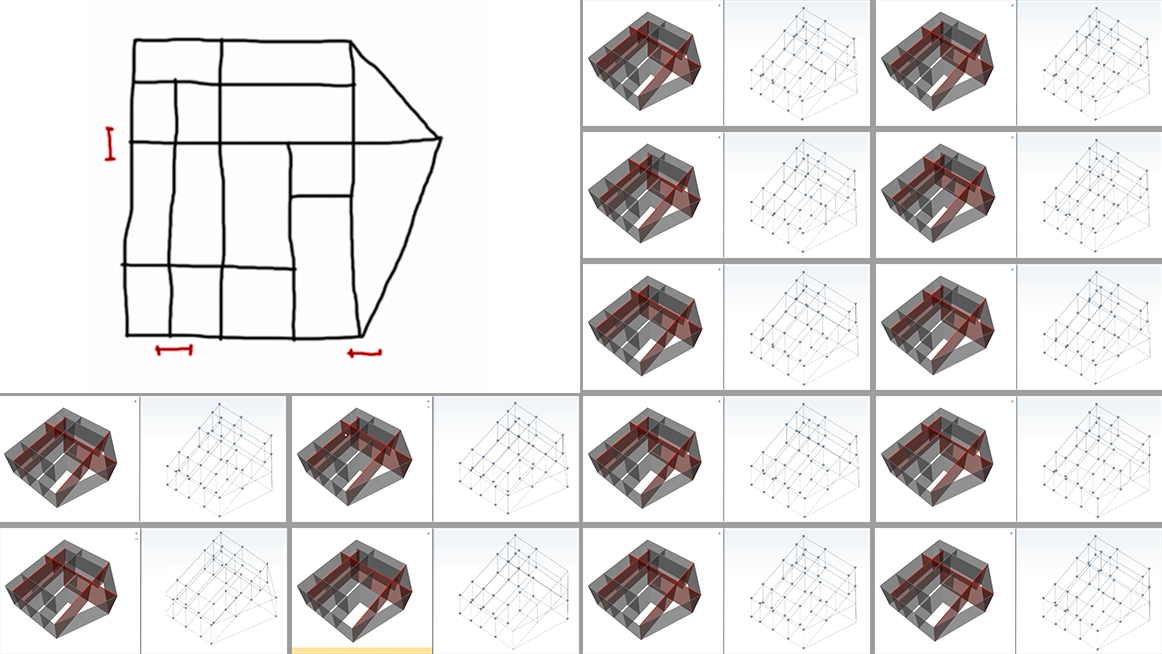}
  \caption{Generated deign solutions}~\label{fig:genDesign}
\end{figure*}

Transformation is assigned to the nodes of the parametric graphs, which each represent an of the corner coordinates of the target wall elements.  This would modify all lines connected to the transformed node. However, to avoid distortion of the orthogonal nature of the plans, we merge colinear paths which connect to each other with a mutual node and share the same direction vector. Next, we identify the array of nodes that are located on the collinear lines. After applying transformations to the connected line node array, we construct new polylines from each node array. This would result in a fully automated parametric model that takes transformation vectors and connected line indexes as an input and outputs a new floorplan layout without producing undesired gaps and floorplan voids.

Various parametric functions are also accessible using the corresponding components in the Parametrizer plugin-in for Dynamo, giving the user the ability to customize the parametrization process in a controlled setting. Grouping can be done based on different criteria (connected lines, adjacent nodes, etc.), which can also be modified by the user of the system to achieve maximum correlations with the design objectives. Furthermore, the parametric model generated in Dynamo can be reconstructed in other analysis tools, such as Autodesk Robot, Radiance \cite{ward1994radiance} or EnergyPlus \cite{crawley2001energyplus}, allowing bi-directional data interaction between the parametric and analysis model. Such practice can result in a faster feedback and iteration process from each sketching attempt as simulated results are updated upon each update of floorplan drawing. 

\subsection{3.3	Design constraint integration}

While the floorplan elements of the parametric model generated in the previous steps are constrained only by its geometric properties, the designer can additionality identify specific optimization variables narrow the solution search to certain geometric modifications. In this regard, we allow the user to specify the variable elements and their corresponding translation constraints by drawing I shaped annotations. The position and orientation of these annotations would allow the system to identify the specific parametric group and allocate a certain translation range based on the length of the annotation. 

Moreover, the user can manually access the target variables with implementing specific components of Parametrizer plugin-in, to modify or visualize the elements using the native Dynamo number slider. This would allow additional design control over the generated parametric model by the user, offering a parallel manual workflow to the automatic procedure. 
\subsection{Generation and Optimization }
We integrate our system with Project Refinery, a NSGA-II optimization tool in Dynamo, to execute building performance optimization tasks. By defining SketchOpt components as input parameters, and assigning custom design goals, such as topological outcomes or performance simulations for the solver. The system can search and evaluate the fitness of each generated model. The parametric constraints specified by sketch-based annotation are also automatically integrated into the workflow. Figure \ref{fig:genDesign} shows examples of the solution space which can also be explored using Project Refinery’s visualization tools. 
\section{Case Study}

We illustrate the use case of our proposed system for a structural optimization case study. Given a building floorplan sketch, the goal of the study is to find the optimized position of the annotated target walls to minimize the total stress and torsion of a simulated structural system. We integrate our system with Autodesk Robot, a structural analysis tool. For simplicity, we define each analytical column by extruding the intersection points of the floorplan walls and construct the beams on the top and bottom edge of the wall.  Figure 3 \ref{fig:sketchPorc} multiple stages of the sketching process drawn in a sketching tool (in our case Adobe Photoshop), and its automatic conversion to a parametric model. Due to the fast processing time of performance metric, we can also observe real-time updates in our structural analysis model. 

Furthermore, the user defines three wall axes to explore how the translation of these walls can achieve the structural objective function. This process is achieved by adding simple $I$ shaped annotations to the floorplan sketch, adjacent to the wall axes (Figure \ref{fig:applyParam}. The system will identify all the corresponding intersection nodes included in the target wall axes. The length of annotation also defines the parametric constraints the walls contain in their solution spaces. Finally, by using Project Refinery’s interface different optimization settings can be assigned to allow the user to control the optimization accuracy and its calculation time. This would result in the customized visualization of the optimization results in Project Refinery with the ability to explore the design solutions in real-time (Figure 5). By hovering on each solution point, the parametric model in Dynamo and structural analysis model in Robot would update allowing the user-designer to investigate different properties of the generated results.

\section{Conclusion}

In this study, we introduce a sketch-based generative design system for building performance optimization tasks. Targeted for amateur users, SketchOpt can be applied in the early stages of the design process, to facilitate architects and designers with limited programming experience in performance-based decision making. The system can assist in form-finding and layout planning in a fast and iterative process. Our next steps for further developing this system is generating optimized multi-level floorplans and 3D building volumes using conceptual sketches of the building. Semantic recognition of furniture layout sketches can also be considered possible ways of improvement. Moreover, our system can be improved with integrating with robust deep learning architecture for faster processing and solution generation procedures.


\begin{acks}
This study was funded and supported by Autodesk, Inc. We would like to thank our colleagues at Autodesk Research and also Autodesk's Architecture, Consturction and Engineering (AEC) Generative Design team for their technical support and feedback. Part of this work is currently patent pending under the U.S Patent Application 16/681,591 \cite{bergin2020automated}.

\end{acks}

\bibliographystyle{ACM-Reference-Format}
\bibliography{sample-base}

\appendix

\end{document}